\documentclass[aps,prl,twocolumn,floatfix,superscriptaddress]{revtex4-1}
\usepackage{graphicx,bm}
\usepackage{times}
\usepackage{amsmath,amssymb}
\usepackage{color}
\usepackage[normalem]{ulem}

\newcommand{\beginsupplement}{%
	\setcounter{table}{0}
	\renewcommand{\thetable}{S\arabic{table}}%
	\setcounter{figure}{0}
	\renewcommand{\thefigure}{S\arabic{figure}}%
}

\begin{document}

\title{Many-body multi-valuedness of particle-current variance in closed and open cold-atom systems}

\author{Mekena Metcalf}
\affiliation{School of Natural Sciences, University of California, Merced, Merced, CA 95343, USA.}
\author{Chen-Yen Lai}
\affiliation{Theoretical Division, Los Alamos National Laboratory, Los Alamos, New Mexico 87545, USA}
\affiliation{Center for Integrated Nanotechnologies, Los Alamos National Laboratory, Los Alamos, New Mexico 87545, USA}
\author{Massimiliano Di Ventra}
\affiliation{Department of Physics, University of California, San Diego, La Jolla, CA 92093, USA.}
\author{Chih-Chun Chien}
\email{cchien5@ucmerced.edu}
\affiliation{School of Natural Sciences, University of California, Merced, Merced, CA 95343, USA.}

\date{\today}

\begin{abstract}
The quantum variance of an observable is a fundamental quantity in quantum mechanics, and the variance provides additional information other than the average itself. 
By examining the relation between the particle-current variance $(\delta J)^2$ and the average current $J$ in both closed and open 
interacting fermionic systems, we show the emergence of a multi-valued Lissajous curve between $\delta J$ and $J$ due to interactions. As a closed system we considered the persistent current in a benzene-like lattice enclosing an effective magnetic flux and solved it by exact diagonalization. For the open system, the steady-state current flowing through a few lattice sites coupled to two particle reservoirs was investigated using a Lindblad equation.
In both cases, interactions open a loop and change the topology of the corresponding $\delta J$-$J$ Lissajous curve, showing that this effect is model-independent.
We finally discuss how the predicted phenomena can be observed in ultracold atoms, thus offering an alternative way of probing the dynamics of many-body systems.
\end{abstract}


\maketitle


Quantum fluctuations of the current flowing in a system provide more information than the average current itself~\cite{Landauer:cw,DiVentra:2010ks,ClerkRMP10}. This fact has been demonstrated in several experimental and theoretical studies ranging from quantum dots~\cite{Gustavsson06} to nanoscale systems~\cite{Chen07,Na14}, to name a few.
In all those studies, {\it two-time} correlations of current are measured (or calculated) away from the average current, and from their spectrum one can infer the type of physical processes at play~\cite{DiVentra:2010ks}.
On the other hand, equal-time density fluctuations at different spatial locations have been measured in ultracold atoms~\cite{Esteve06}, revealing spatial correlations in quantum gases.

However, one could also study quantum {\it variance} of the current, a property of fundamental importance in quantum mechanics because of the uncertainty principle~\cite{SakuraiQuantum}.
This equal-time, equal-space quantity has been less explored, presumably because of experimental difficulty in measuring it in a current-carrying system.
Emergence of cold atoms~\cite{Pethick:2010gy,ueda2010fundamentals,stoof2008ultracold} as new model systems to study a host of phenomena otherwise difficult to
probe using traditional solid-state materials, makes this transport property readily accessible experimentally~\cite{Chien:2015kc}.
It is then natural to ask what information the variance would reveal, and how that information might be useful in characterizing the many-body dynamics.

In this paper, having in mind cold-atom systems as possible experimental verification of our predictions, we study the quantum variance, $(\delta J)^2$, of current flowing in a fermionic many-body system, and relate this quantity to the average current, $J$.
We consider two experimentally realizable situations: the persistent current of a periodic system and the steady-state limit of the current in an open system. 
The latter case is more amenable to an easier experimental realization in ultracold atoms~\cite{Krinner17,Lebrat18}.

In order to solve the many-body problem exactly (hence beyond mean field), we have considered, as a closed system, a benzene-like ring lattice with a static magnetic flux to sustain a persistent current.
The open system is a triple-site lattice connected to two particle reservoirs with tunable system-reservoir couplings.
Exact diagonalization~\cite{ED_Alg,ED1993} is used to find the closed system many-body ground state, and a Lindblad equation~\cite{weiss2012quantum,Lai:2018vx,Lai:2016kh} is implemented to simulate the time-evolved density matrix for the open system.

In both cases we plot the square root of the quantum variance, or the standard deviation $\delta J$, versus average current $J$ as the interaction strength varies. We find two distinct classes of the $\delta J$-$J$ Lissajous curve~\cite{StaufferBook, LissajousEng, BasicElec}: (1) A one-to-one correspondence between $\delta J$ and $J$ in the absence of interactions except possible isolated points due to energy degeneracy and (2) a loop or more complicated patterns in the presence of interactions.
The Lissajous loop thus establishes a multi-valued relation between $\delta J$ and $J$ for interacting many-body systems.
The similarity of results in both closed and open systems suggests our findings are model-independent.
However, as the interaction vanishes, the Lissajous curve closes abruptly in the closed system case but smoothly in the open system one.
The difference is due to quantum degeneracy of noninteracting fermions in the closed case versus the open one.
Irrespective, these results show that the presence of interactions between the particles can be detected as the degree of {\it multi-valuedness} of the corresponding $\delta J$-$J$ Lissajous curve.
We finally discuss how to verify this many-body effect in cold-atom experiments.

\begin{figure}[t]
	\begin{center}
		\includegraphics[width=\columnwidth]{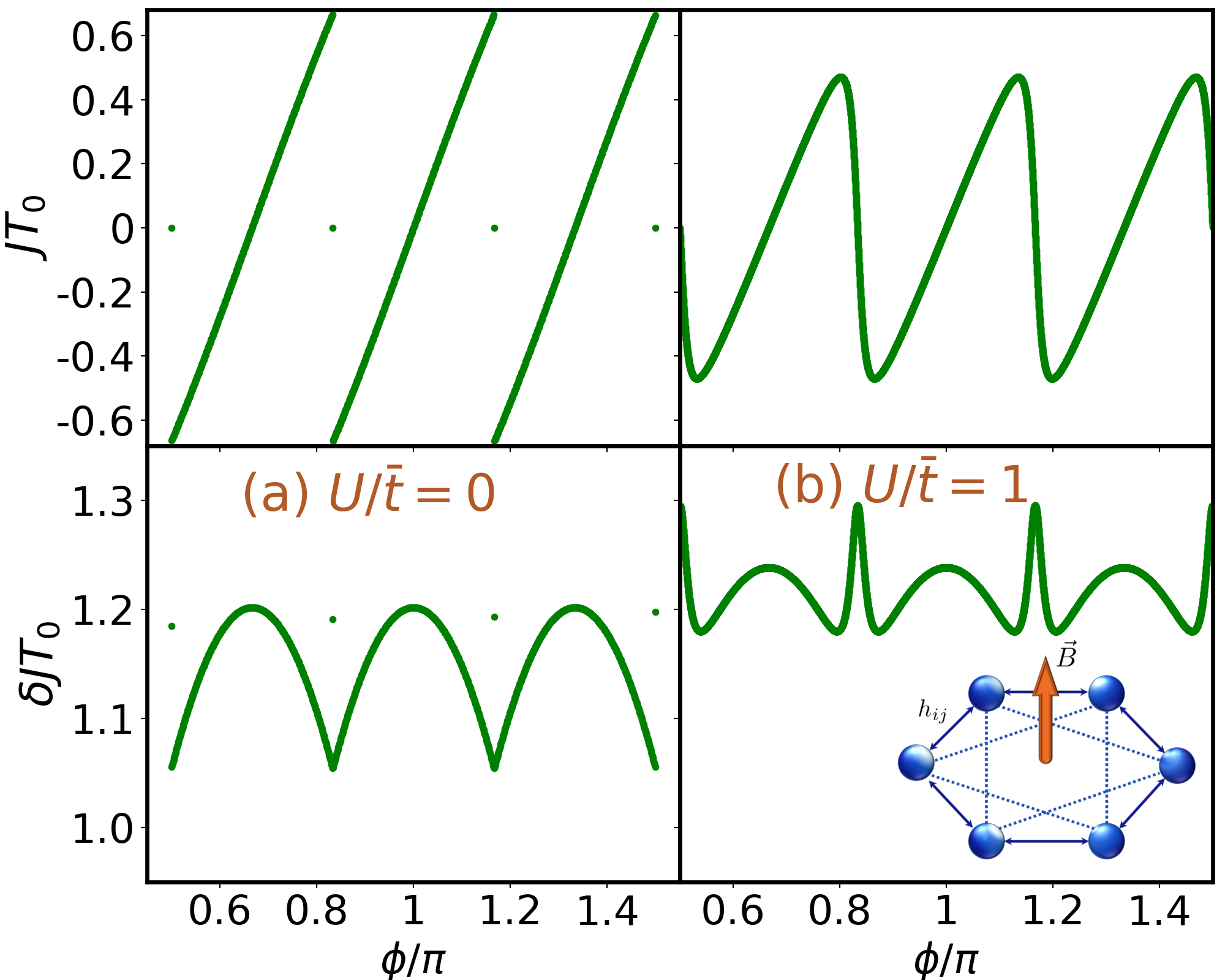}
		\caption{
			A benzene-like lattice with $N_\uparrow \!=\! N_\downarrow \!=\! 3$ fermions threaded by a magnetic field perpendicular to the plane (illustrated in the inset).
			The hopping coefficient from site $i$ to site $j$ is $h_{ij}$.
			The onsite coupling constant is $U$ and the tunneling coefficient is $\bar t$.
			Ground-state persistent currents (top row) and its standard deviation (bottom row) as functions of the Peierls phase $\phi$ for (a) noninteracting and (b)
			$U/\bar{t}\!=\!1$. The time unit $T_0= \hbar /\bar t$ where $\hbar = 1$.
		}
		\label{fig:CQS_NNJdJ}
	\end{center}
\end{figure}

{\it Closed system --} As a model closed system amenable to exact diagonalization, we consider a benzene-like ring lattice with two-component fermions (labeled by the spin $\sigma\!=\uparrow,\downarrow$) hopping between the six sites.
The lattice is half-filled with same number of up-spin and down-spin fermions, $N_\uparrow\!=\!N_\downarrow$.
When a perpendicular, static magnetic field threads the ring, as illustrated in Fig.~\ref{fig:CQS_NNJdJ}(b),
a persistent current in the ring emerges~\cite{HubbardPC_BetheSoln, AkkermanBook, QuantumRingsForBeginners, PhysicsofZero1DNanoSystems}.
The effect of the magnetic flux is included by the Peierls substitution~\cite{PeierlsOrig, PeierlsUCA2012} in the tight-binding approximation, and we model the system by the Fermi-Hubbard model with the Hamiltonian
\begin{equation}\label{eq:H}
	\hat H = \sum_{i\neq j,\sigma}h_{ij} \hat c_{i\sigma}^{\dagger}\hat c_{j\sigma} + U\sum_i \hat n_{i\uparrow}\hat n_{i\downarrow}.
\end{equation}
Here $\hat c_i^{\dagger}$ ($\hat c_i$) denotes the fermion creation (annihilation) operator at site $i$, and $\hat n_i\!=\!\hat c_i^{\dagger}\hat c_i$.
The hopping coefficient between site $i$ and site $j$ is $h_{ij}$ and $h_{ji}\!=\!h_{ij}^*$.
Onsite contact interactions between fermions of opposite spins has the coupling constant $U$.

We first focus on a system with only nearest-neighbor hopping, $h_{i,i+1}\!=\!\bar{t}e^{i\phi}$.
Here, $\bar{t}$ is the tunneling coefficient and $\phi$ is the Peierls phase.
For a uniform magnetic field, $\phi$ is proportional to the magnetic field strength~\cite{PeierlsOrig,AkkermanBook}.
The energy unit is $\bar{t}$ and the time unit is $T_0\!=\!\hbar/\bar{t}$.
We consider the zero-temperature case, where the persistent current is a property of the many-body ground state~\cite{AkkermanBook, QuantumRingsForBeginners}.
For a moderate lattice size, we use the exact diagonalization technique~\cite{ED_Alg,ED1993} to obtain the ground state and excited states along with their energies.

The current operator of the fermions with spin $\sigma$ from site $i$ to site $j$ is
\begin{equation}\label{eq:Jop}
	\hat J_{ij,\sigma} = i(h_{ij}\hat c_{i\sigma}^{\dagger}\hat c_{j\sigma}- h_{ij}^*\hat c_{j\sigma}^{\dagger}\hat c_{i\sigma}).
\end{equation}
At zero temperature, the persistent current is the expectation value with respect to the many-body ground state, $\langle \hat J_{ij,\sigma}\rangle$.
The current variance of a single spin component is
\begin{equation}
	\delta J_{ij,\sigma}^2 = \langle \hat J_{ij,\sigma}^2\rangle -\langle \hat J_{ij,\sigma}\rangle^2.
\end{equation}
By using Eq.~\eqref{eq:Jop} and the density operator of the fermions with spin $\sigma$ on site $i$, $\hat n_{i\sigma}\!=\!\hat c^{\dagger}_{i\sigma} \hat c_{i\sigma}$, we obtain
\begin{equation}\label{eq:J2}
	\langle \hat J_{ij,\sigma}^2\rangle = \big| h_{ij}\big|^2\big[ \langle \hat n_{i\sigma} \rangle + \langle \hat n_{j\sigma} \rangle - 2\langle \hat n_{i\sigma} \hat n_{j\sigma}\rangle\big].
\end{equation}
Therefore, the current variance reflects the density-density correlations between the sites across which the current flows. Since the system is translationally invariant on a lattice, we will drop the subscript $ij$.

The total current is the sum over the spin components
\begin{equation}
\langle \hat J\rangle = \langle \hat J_{\uparrow}\rangle + \langle \hat J_{\downarrow}\rangle,
\end{equation}
and since we work at half-filling, $\langle \hat J_{\uparrow}\rangle\!=\!\langle \hat J_{\downarrow}\rangle$ and $\langle \hat J_{\uparrow}^2\rangle\!=\!\langle \hat J_{\downarrow}^2\rangle$. The total current variance is then
\begin{equation}
\delta J^2= 2  \langle \hat J_{\uparrow}^2\rangle + 2\langle \hat J_{\uparrow}\hat J_{\downarrow}\rangle - 4\langle \hat J_{\uparrow}\rangle^2.
\end{equation}
The cross-component current correlation is
$\langle \hat J_{\uparrow}\hat J_{\downarrow}\rangle \!=\!  -(\langle h_{ij}^2 \hat c_{i\uparrow}^{\dagger}\hat c_{j\uparrow}\hat c_{i\downarrow}^{\dagger}\hat c_{j\downarrow}\rangle \!-\! \langle |h_{ij}|^2 \hat c_{i\uparrow}^{\dagger}\hat c_{j\uparrow}\hat c_{j\downarrow}^{\dagger}\hat c_{i\downarrow}\rangle \!-\! \langle |h_{ij}|^2 \hat c_{j\uparrow}^{\dagger}\hat c_{i\uparrow}\hat c_{i\downarrow}^{\dagger}\hat c_{j\downarrow}\rangle \!+\! \langle {h^*_{ij}}^2 \hat c_{j\uparrow}^{\dagger}\hat c_{i\uparrow}\hat c_{j\downarrow}^{\dagger}\hat c_{i\downarrow}\rangle)$.
As expected, in the presence of interactions, the total current variance is not generally a simple sum of the current variance from each spin component, i.e., $\delta J^2\!\neq\!\sum_{\sigma} \delta J_{ij,\sigma}^2$. Only in the noninteracting case with an equal population of both species, the Wick decomposition leads to an equality between the total current variance and the sum of the current variance from the two spins.

The persistent current $J$ and its standard deviation $\delta J$ of non-interacting fermions in the benzene-like lattice are shown in Fig.~\ref{fig:CQS_NNJdJ}(a).
They exhibit periodic structure as the Peierls phase $\phi$ increases. We remark that each value of $\phi$ corresponds to a static magnetic flux and the persistent current is an equilibrium property~\cite{AkkermanBook}.
Interestingly, there are discontinuities in both $J$ and $\delta J$, as shown by the isolated points in Fig.~\ref{fig:CQS_NNJdJ}(a).
These discontinuities are due to level crossings in the energy spectrum, which are known in the study of persistent currents in a ring~\cite{HubbardPC_BetheSoln, AkkermanBook, QuantumRingsForBeginners}.
(See the Supplemental Information (SI) for details of the level crossing.)
At a level-crossing point, the values of $J$ and $\delta J$ are determined by assigning each degenerate state equal statistical weight, so the result is consistent with the zero-temperature limit~\cite{AkkermanBook}.
Similar discontinuities have also been observed in the superfluid velocity and its square for clean superconductors in the Little-Parks experiment (see, e.g., Ref.~\cite{Tinkham_SCbook}).

\begin{figure}[t]
	\begin{center}
		\includegraphics[width=\columnwidth]{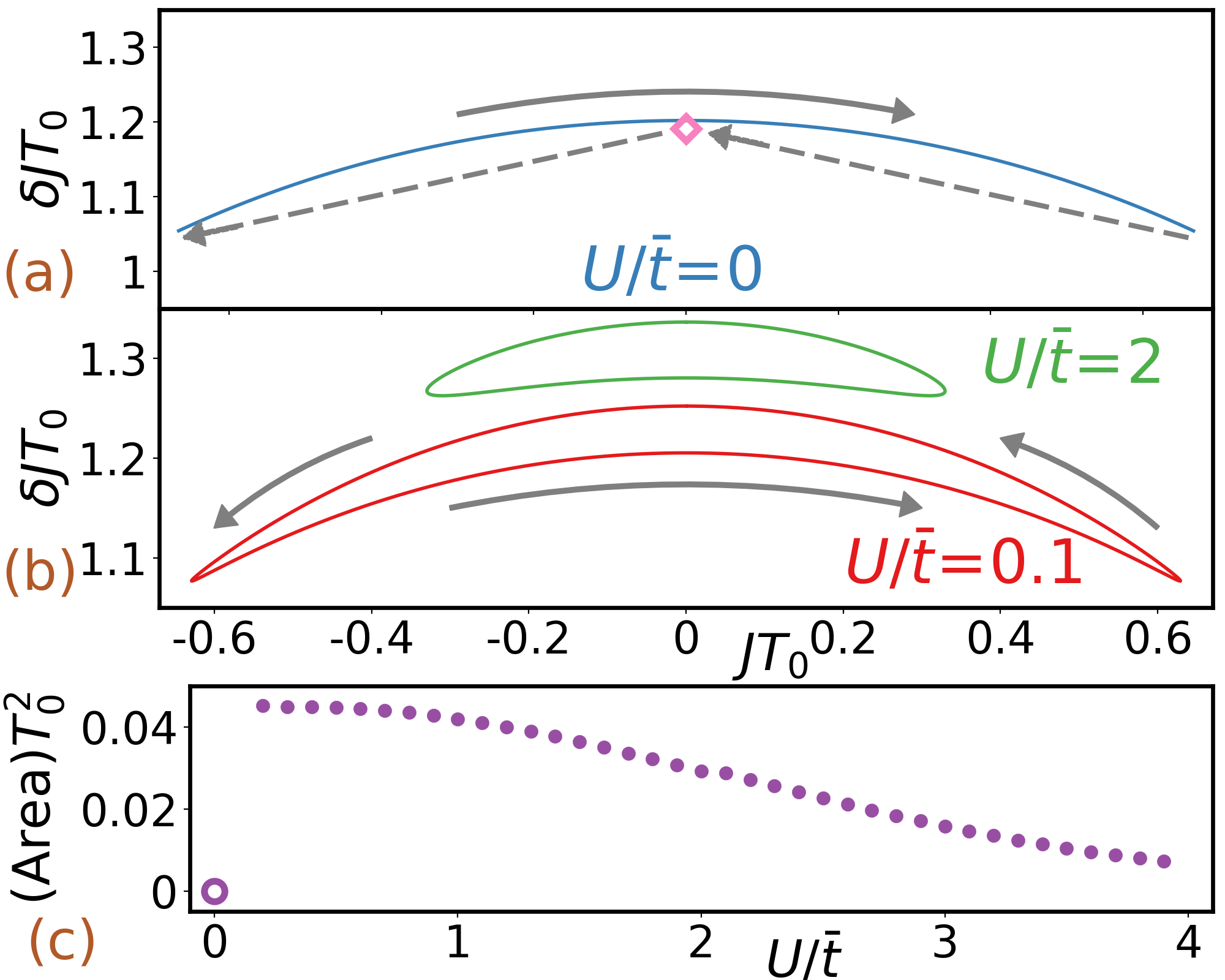}
		\caption{
			Ground-state Lissajous curve of $\delta J$ vs. $J$ in the range $\pi/2 \leq \phi \leq 5\pi/6$ for (a) $U/\bar{t} \!=\! 0$, (b) $U/\bar{t} \!=\! 0.1$ (red) and $U/\bar{t} \!=\! 2$ (green).
			The gray arrows indicate the direction of increasing magnetic flux.
			The red rhombus in (a) is an isolated point due to the degeneracy point, and the dashed arrows indicate the discontinuous jumps. The interaction turns the degeneracy point into avoided crossing, and the isolated point in (a) becomes the continuous upper part of the Lissajous curves shown in (b).
			(d) Area of the loop over one period.
			The area is ill-defined at $U/\bar{t} \!=\! 0$ and is denoted by an open circle at zero.
			The area jumps to a non-zero value abruptly when the interaction is present.
		}
		\label{Fig:HystArea_Fluc}
	\end{center}
\end{figure}

In the presence of the onsite interaction, the persistent current and its standard deviation become continuous curves as shown in Fig.~\ref{fig:CQS_NNJdJ}(b),
where the result follows from exact diagonalization of the Hamiltonian~\eqref{eq:H}.
This is because the interaction turns the level crossings into avoided crossings. (See the SI for details.)
The current can then be viewed as a continuous, periodic function of $\phi$.
The same features can also be observed in $\delta J$.
Importantly, the skewed dome-shape curve of $J$ vs. $\phi$ allows one to find two different values of $\phi$ giving rise to the same value of $J$ within the same period. Importantly, this feature is absent in the noninteracting case.

To elucidate the relation between $\delta J$ and $J$, we use the fact that given two continuous, periodic functions $\alpha(\phi)$ and $\beta(\phi)$, one can form a Lissajous curve~\cite{StaufferBook, LissajousEng, BasicElec} by plotting $\beta(\phi)$ against $\alpha(\phi)$.
Figure~\ref{Fig:HystArea_Fluc} shows the Lissajous plots of $\delta J$ vs. $J$ for the noninteracting and interacting cases.
It is clear that Lissajous curves for the interacting system shows hysteresis, similarly to the hysteresis loop in magnetization~\cite{StaufferBook, BertottiHystBook, MayergoyzHyst}.
On the other hand, the noninteracting system exhibits only a single curve from the continuous part and an isolated point from the discontinuous jumps of both the current and its variance at the degenerate point, as shown in Fig.~\ref{Fig:HystArea_Fluc}(a).
This implies that the standard deviation of current changes continuously along the curves shown in Fig.~\ref{Fig:HystArea_Fluc}(b) for the interacting case but abruptly jumps to the isolated point in the noninteracting case shown in Fig.~\ref{Fig:HystArea_Fluc}(a).
Thus, there is a topological distinction between the Lissajous curves of the non-interacting and interacting systems because the set of points on the $\delta J$ vs. $J$ plot is simply-connected in the interacting case and multiply-connected, due to the isolated point, in the noninteracting case.

To further quantify the topological difference between the the Lissajous curves of the noninteracting case and the interacting case, we calculate the area enclosed by the loop of the $\delta J$ vs. $J$ curve and show its dependence on $U$ in Fig.~\ref{Fig:HystArea_Fluc}(c).
As $U\!\rightarrow\!0$, the area of the loop approaches asymptotically a constant value.
However, there is no area in the noninteracting case because the curve does not form a loop due to the degenerate points.
The transition from a non-zero area loop to a curve plus an isolated point is therefore very sharp.
We have indeed verified that a non-zero loop can still be observed down to $U/\bar{t}\!=\!0.01$.
Incidentally, the half-filled lattice approaches the Mott insulating phase as the repulsive interactions are increased~\cite{Mahan_CMN, Lieb1DHubbard}.
As a consequence, both the current and current variance are suppressed and the area of the $\delta J$ vs $J$ loops decreases with increased interactions, as shown in Fig.~\ref{Fig:HystArea_Fluc}(c).

We finally remark that the Lissajous curves and the interaction-induced change of the topology of the $\delta J$ vs. $J$ plot are still observable in the presence of weak next nearest neighbor (NNN) hopping (see the SI).
More complicated structures in the $\delta J$ vs $J$ plot, however, can be induced by adding strong NNN hopping.
In the presence of moderate, attractive interactions ($U\!<\!0$), the Lissajous curves are multivalued but do not form closed loops due to degeneracy points in the energy spectrum (see the SI). Adding more lattice sites to the closed system does not change the conclusions, and we have checked the results up to $L=8$ sites (See the SI). 
The $\phi$-period of $J$ and $\delta J$, however, shrinks as the lattice number increases, as well as the amplitude of the current.
We have also checked the necessity of interactions for the loop formation in the Lissajous curve by verifying the absence of any loop away from half-filling for the non-interacting case (see the SI).

\begin{figure}[b]
  \begin{center}
    \includegraphics[width=\columnwidth]{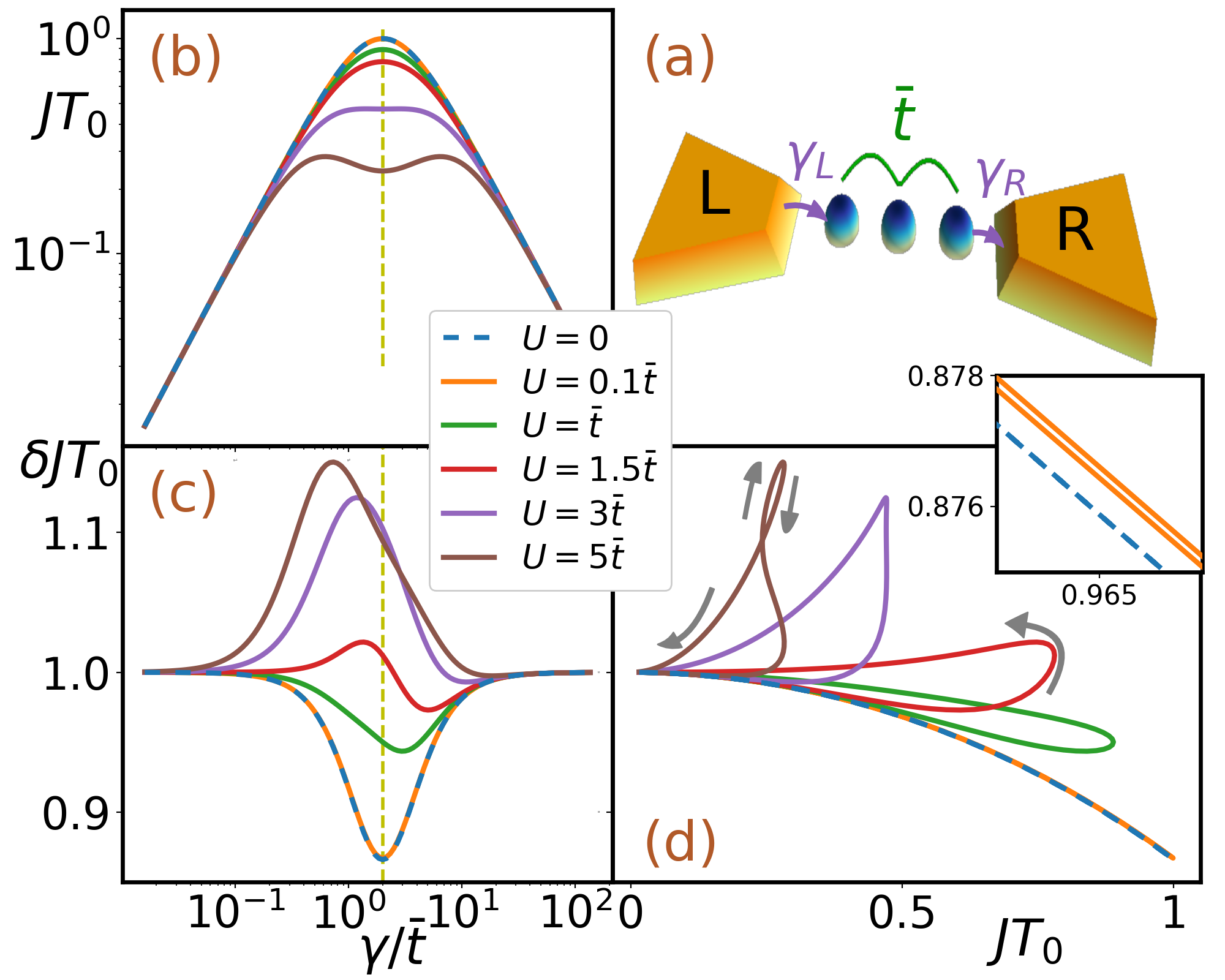}
		\caption{
			(a) Schematic plot of a triple-site lattice connected to a source (left) and drain (right). The hopping coefficient between the lattice sites is $\bar{t}$, the system-reservoir couplings are $\gamma_{L,R}$, and the onsite interaction has the coupling constant $U$. The open system is described by the quantum master equation~\eqref{eq:lindblad}.
			(b) Current $J$ and (c) standard deviation of the current $\delta J$ versus $\gamma\!=\!\gamma_{L}\!=\!\gamma_{R}$ for selected values of $U/\bar{t}$ in the triple-site open system.
			When $U\!=\!0$, the current and current variance are symmetric about the yellow dashed line ($\gamma\!=\!2\overline{t}$).
			When $U/\bar{t}\!>\!0$, only the current remains symmetric while the standard deviation of current becomes asymmetric.
			(d) $\delta J$ versus $J$.
			When $U\!=\!0$ (dashed line), there is a one-to-one correspondence between the current and its standard deviation.
			In contrast, a closed loop and multi-valuedness are observed for all $U\!>\!0$.
			The arrows indicate the direction of increasing $\gamma$.
			The inset shows a detailed comparison of $\delta J$ vs. $J$ for $U/\bar{t}=0$ and $U/\bar{t}\!=\!0.1$.
		}
		\label{fig:OQS_3QDs_U}
	\end{center}
\end{figure}

{\it Open systems --}
In order to show that the phenomenon we have just described is model-independent and accessible by available cold-atom technology, we demonstrate its emergence in a triple-site lattice system coupled to two external particle reservoirs via the first and last sites, as illustrated in Fig.~\ref{fig:OQS_3QDs_U}(a).
This type of system is more easily realizable in cold-atom experiments~\cite{Krinner17,Lebrat18}.
The Hamiltonian $\mathcal{H}_{\text{S}}$ of the triple-site lattice is given by Eq.~\eqref{eq:H} with $i\!=\!1,2,3$, $h_{ij}\!=\!\bar{t}$.
To model the system coupled to reservoirs, we follow an open-system approach~\cite{weiss2012quantum,Lai:2018vx} and monitor the dynamics by a quantum master equation describing the time evolution of the density matrix.
The left (right) reservoir acts as a particle source (drain) which pumps (takes) particles into (out of) the three sites with coupling $\gamma_{L}$ ($\gamma_R$).

The dynamics in the open-system approach is described by the Lindblad equation~\cite{breuer2007theory,datta2005quantum,DiVentra:2010ks,weiss2012quantum,Lai:2018vx}:
\begin{eqnarray}\label{eq:lindblad}
	\frac{d}{dT}\hat \rho&=&i\left[\hat \rho,\mathcal{\hat H}_{\text{S}}\right]
	+\gamma_L\sum_{\sigma}\left(\hat c^\dagger_{1\sigma}\hat \rho \hat c_{1\sigma}-\frac{1}{2}\{\hat c_{1\sigma}\hat c^\dagger_{1\sigma},\hat \rho\}\right) \nonumber \\
	& &+\gamma_R\sum_\sigma\left(\hat c_{l\sigma}\hat \rho \hat c^\dagger_{l\sigma}-\frac{1}{2}\{\hat c^\dagger_{l\sigma}\hat c_{l\sigma},\hat \rho\}\right),
\end{eqnarray}
where $\hat \rho$ is the density matrix of the three sites and $l$ is the rightmost site. The assumptions behind the derivation of the Lindblad equation and the details of its numerical solution are summarized in the SI.

Once the time-evolved density matrix is found, the current and current variance can be determined by ($\hat{O}$ is replaced by the corresponding operator)
\begin{equation}
	\langle \hat{O} \rangle = \text{Tr}(\hat \rho\hat{O}) = \sum_{mn}\langle m\vert \hat{O} \vert n\rangle\hat \rho_{nm},
\end{equation}
where Tr denotes the trace and \{$\vert m\rangle$\} is the set of the Fock-space basis kets from all possible particle-number sectors.
In the following, we choose $\gamma_L\!=\!\gamma_R=\gamma$.
Moreover, we focus on the steady state, $d\hat \rho/dT\!=\!0$ in the long-time limit $T\!\rightarrow\!\infty$, where a steady-state current can be identified.

Unlike the closed system where the current and current variance exhibit periodic behavior when the system is driven by an enclosed magnetic flux, the steady-state current of the open system is driven by the two reservoirs and does not have any periodic property.
Nevertheless, as shown in Fig.~\ref{fig:OQS_3QDs_U}(b), the steady-state current through the triple-site lattice exhibits a maximum as $\gamma$ varies.
Hence, there can be two different values of $\gamma$ giving rise to the same value of current $J$.
Furthermore, it has been shown~\cite{Gruss:2016js} that the current exhibits $\gamma$- and $1/\gamma$- dependence in the small and large $\gamma$ regimes.
Thus, the current is symmetric on the log-log plot about $\gamma\!=\!2\overline{t}$, and we found the currents remain symmetric when the interactions are finite.
In contrast, the standard deviation of current, $\delta J$, shown in Fig.~\ref{fig:OQS_3QDs_U}(c) is symmetric about $\gamma\!=\!2\overline{t}$ on the semi-log plot only for the non-interacting fermions.
When $U\!>\!0$, $\delta J$ is asymmetric as $\gamma$ varies.

Therefore, by plotting the current vs. its standard deviation as shown in Fig.~\ref{fig:OQS_3QDs_U}(d), one can check if there are multi-values of $\delta J$ for the same value of $J$.
For the noninteracting case, we found a one-to-one correspondence between $J$ and $\delta J$ following their symmetric curves in Figs.~\ref{fig:OQS_3QDs_U}(b) and (c).
However, when $U/\bar{t}>0$, one can observe loop structures as shown in Fig.~\ref{fig:OQS_3QDs_U}(d).
The loops indicate again multi-valuedness of $\delta J$ vs. $J$.
When the interaction decreases in the open system, the loop in the $\delta J$ vs $J$ plot shrinks accordingly, see Fig.~\ref{fig:OQS_3QDs_U}(d). In contrast to the closed-system case, there is no level crossing in the open system and the transition to the zero-area curve is smooth as the interaction vanishes. Therefore, degeneracy points are not essential in the change of the topology of the Lissajous curve as the interactions are turned on.

Note though, in the strongly-interacting regime the open system can exhibit more complicated loop structures as shown in Fig.~\ref{fig:OQS_3QDs_U}(d), and there can be multiple nodes in the loops signaling another topology change of the Lissajous curve.
In the closed system, only simple loops appear with nearest-neighbor hopping.
Interestingly, the open system with only two sites also exhibits multi-valuedness of $\delta J$ vs. $J$ for both noninteracting and interacting fermions, but there is no loop structure in the $\delta J$ vs. $J$ plot. The difference comes from the fact that both sites in the double-site systems are coupled to the reservoirs, so the system is fully controlled by the reservoirs and does not exhibit intrinsic behavior from the sites. (See the SI for more details.)

{\it Experimental implications --}
As anticipated, the phenomena discussed here may be observed experimentally using ultracold atoms in engineered optical potentials.
For example, the benzene-like lattice may be realized by using atom-by-atom assembly with optical tweezers~\cite{Barredo_AtomAssemb, Endres_AtomAssemb,RegalRev} or painting potentials~\cite{PaintPot2009}.
The magnetic field may be simulated by artificial gauge fields from light-atom interactions, and the Peierls phase has been demonstrated~\cite{PeierlsUCA2012, Lin_MagGauge, SpielmanGaugeRev}.
Open systems with a few lattice sites coupled to two reservoirs have recently been realized by connecting an optical lattice to atom reservoirs~\cite{Lebrat18}.
Importantly, the quantum coherence in ultracold atoms is robust because the system is virtually decoupled from the outside environment.
Direct measurements of the current may be achieved by utilizing auxiliary light-atom coupling~\cite{CurrentMeasurement}, and the current standard deviation may be obtained from the variance of the current from an ensemble measurement.
Cooling the system is important because thermal fluctuations could complicate the interpretation of the result.

{\it Conclusions --}
We have shown that the Lissajous curve of the particle-current and its standard deviation exhibits a loop structure in the presence of interactions for both closed and open fermionic systems.
The loop area is finite only when interactions are present, and the loop structures are robust against size of the system, particle filling, and weak next-nearest-neighbor hopping. Degeneracy points of noninteracting or attractive interacting systems, instead, lead to discontinuities of the current and its variance.
The effects we report here are within experimental reach for ultracold atoms in engineered optical potentials and the $\delta J$ vs. $J$ Lissajous curve demonstrates interesting relations between quantum expectation values and their variance.
The multi-valuedness of the averaged current vs. its quantum variance can then be used as an alternative way to discern interactions in these many-body systems. We note also that, due to the different spin-statistics, the particle-current quantum variance of bosons may also exhibit interesting behavior. However, their larger Fock space requires more demanding computations and we defer the study for future investigations.

\acknowledgments{We thank George Chapline, David Weld, and Kevin Wright for stimulating discussions and Leland Ellison for critical comments. C.Y.L. acknowledges the support from the Center for Integrated Nanotechnologies, a U.S. Department of Energy, Office of Basic Energy Science user facility.}


\pagebreak

\begin{widetext}
	\begin{center}
		\textbf{Supplemental Information: Many-body multi-valuedness of the particle-current variance in closed and open cold-atom systems}
	\end{center}
\end{widetext}
\beginsupplement

\section{Additional information for closed systems}
	\subsection{Level crossing and avoided crossing}
	\begin{figure}[b]
		\begin{center}
			\includegraphics[width=\columnwidth]{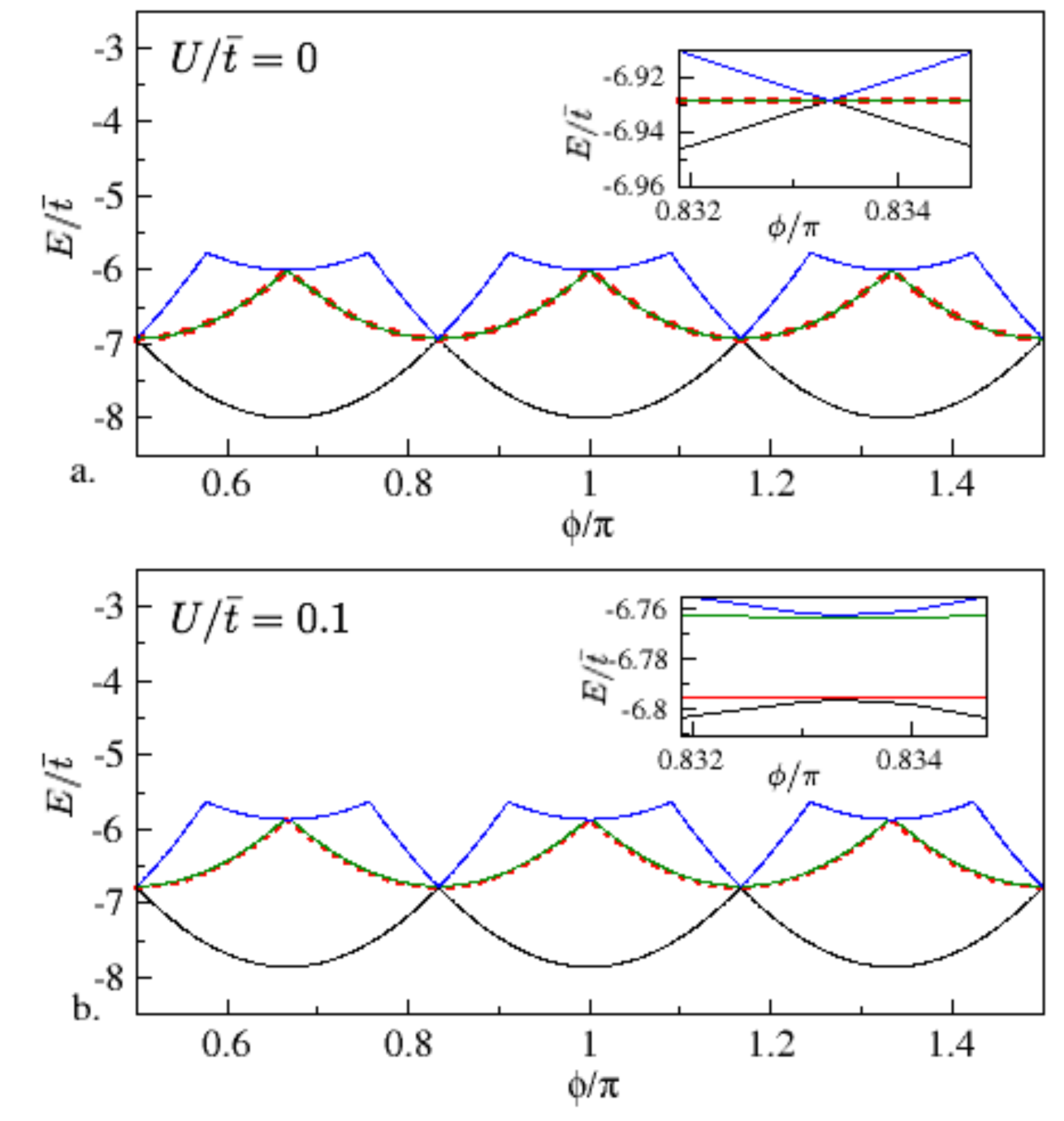}
			\caption{Energies of the lowest four states as a function of the Peierls phase $\phi$ of a half-filled 6-site lattice loaded with (a) non-interacting and (b) weakly interacting ($U/\bar{t} = 0.1$) fermions. The inset of (a) show magnified views around the degeneracy point of the noninteracting case. The inset of (b) shows the level crossings of the noninteracting case are turned into avoided level crossing in the presence of interactions. }
			\label{Fig:Energy}
		\end{center}
	\end{figure}
	A benzene-like lattice with noninteracting fermions and (next-)nearest neighbor hopping has degenerate points periodic in the magnetic flux, as shown in Fig~\ref{Fig:Energy}. Four states are degenerate at those degenerate points. In the presence of interactions, the degeneracy is lifted. Thus, the level crossing becomes avoided crossing.
	Fig~\ref{Fig:Energy}(b) shows the avoided crossing when $U/\bar{t}=0.1$. Since the energy curves are smooth around the avoided crossing, the current and current variance become continuous functions of the Peierls phase.
	
	\subsection{From magnetic field flux to Peierls phase}
	\begin{figure}[b]
		\begin{center}
			\includegraphics[width=3.5in]{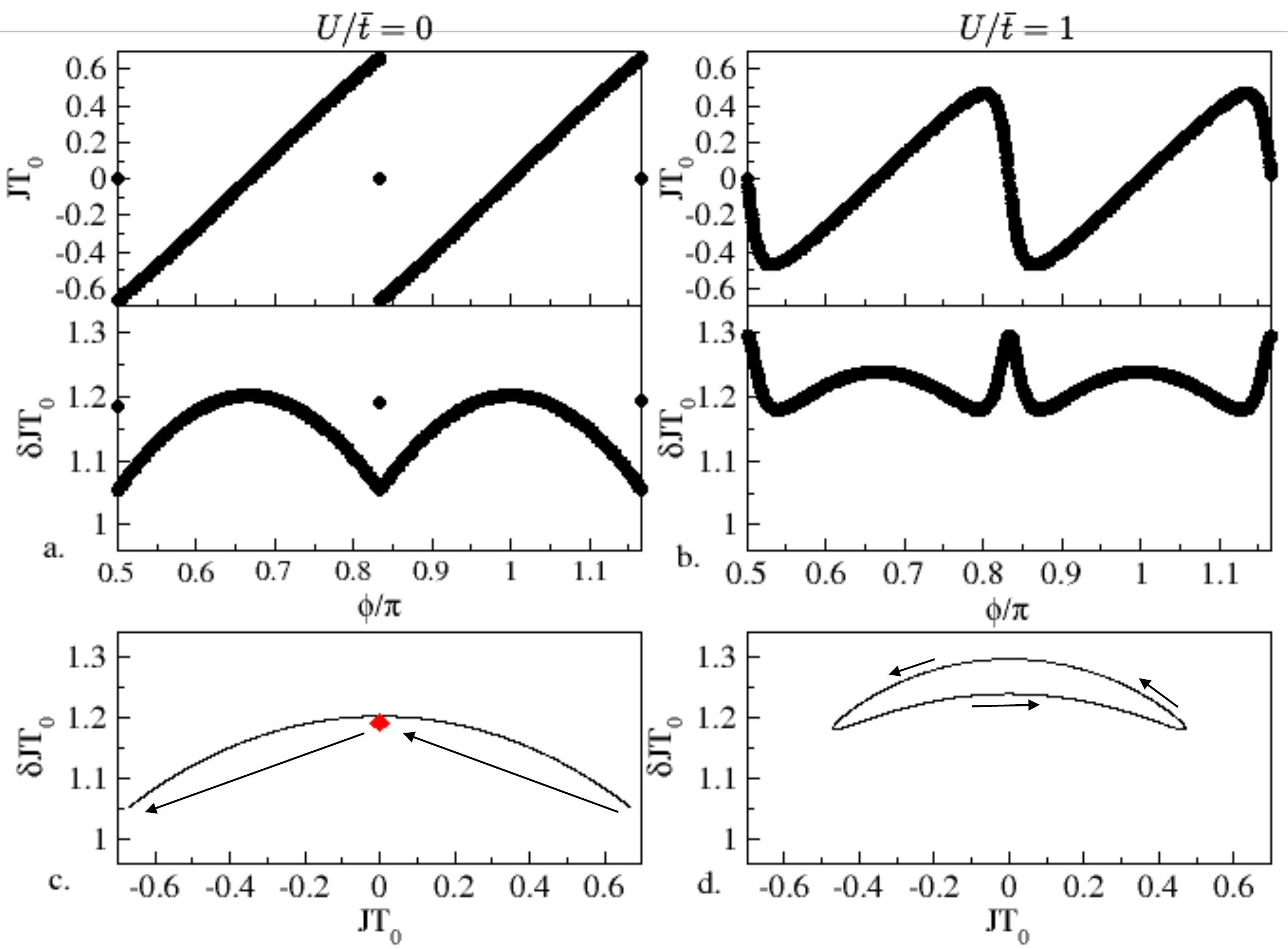}
			\caption{ (a) and (b) The current $J$ and its standard deviation $\delta J$ as functions of the Peierls phase $\phi$. (c) and (d) $\delta J$ vs. $J$ Lissajou curve in the range $\pi/2 \leq \phi \leq 5\pi/6$ for (a) and (c) $U/\bar{t} = 0$ and (b) and (d) $U/\bar{t} = 1$. The benzene-like lattice has both nearest-neighbor and next-nearest neighbor hopping with $\bar{t'} = \bar{t}/5$. The red rhombus in (c) indicate the discontinuity from the degeneracy point. The arrows show how the Lissajou curves evolve as $\phi$ increases.}
			\label{Fig:Curr&Hyst_NNN5}
		\end{center}
	\end{figure}
	
	To find the gauge-invariant expression of the Peierls phase, we associate the magnetic flux with the vector potential on each link of the lattice. The magnetic flux is
	\begin{equation}
	\Phi = \oint B\cdot dS= \oint A \cdot dl.
	\end{equation}
	Here $B$ is the magnetic field perpendicular to the ring lattice, $dS$ is the surface element, $A$ is the vector potential, and $dl$ is the line element connecting the sites.
	
	For cold-atoms, the vector potential can be induced by light-atom interactions~\cite{Lin_MagGauge}. The Peierls phase $\phi=\oint A dl$ of one link is $\phi = \Phi/N$ if we assume the vector potential is uniform. Here $N$ is the number of links enclosing the flux. If we assume a uniform $B$ field, a straightforward analysis shows that we should choose $h_{ij}=\bar{t}e^{i\phi}$ for the nearest-neighbor (NN) link and $h_{ij} = \bar{t}'e^{i\phi}$ for the next-nearest-neighbor (NNN) link. Here $\bar{t}$ and $\bar{t}'$ are the tunneling coefficients in the absence of the field. Analyzing ring geometry and total flux through the enclosed area, the Peierls phase is the same for NN and NNN links.
	
	\begin{figure}[t]
		\begin{center}
			\includegraphics[width=3.5in]{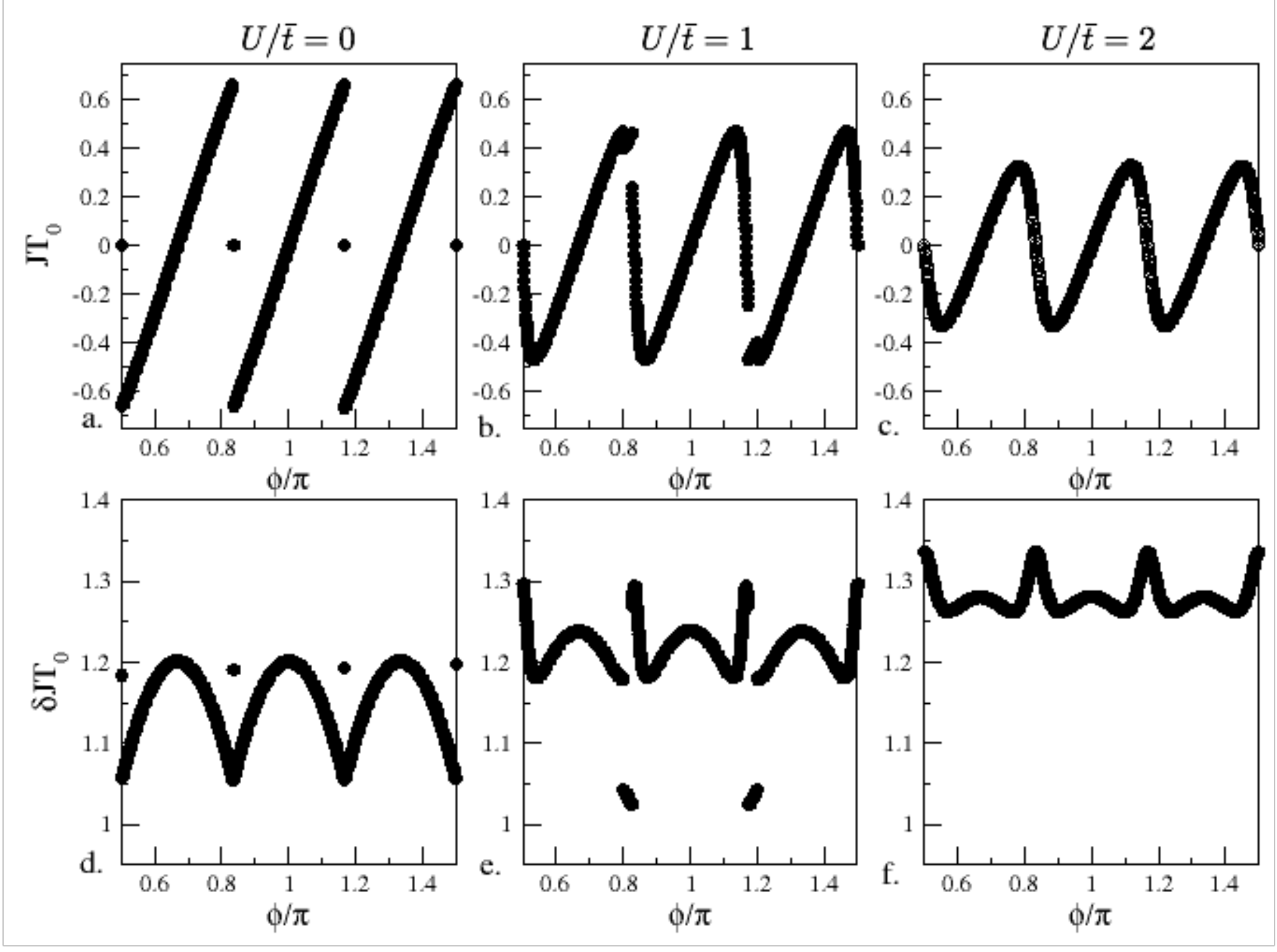}
			\caption{ (a-c) Ground-state persistent currents and (d-f) corresponding current fluctuations as functions of the Peierls phase $\phi$ for $U/\bar{t} = 0$, $U/\bar{t} = 1$, and $U/\bar{t} = 2$ with both the NN and NNN hopping. Here $\bar{t'} = \bar{t}/2$. }
			\label{Fig:Curr&Hyst_NNN2}
		\end{center}
	\end{figure}
	
	\begin{figure}
		\includegraphics[width=3.5in]{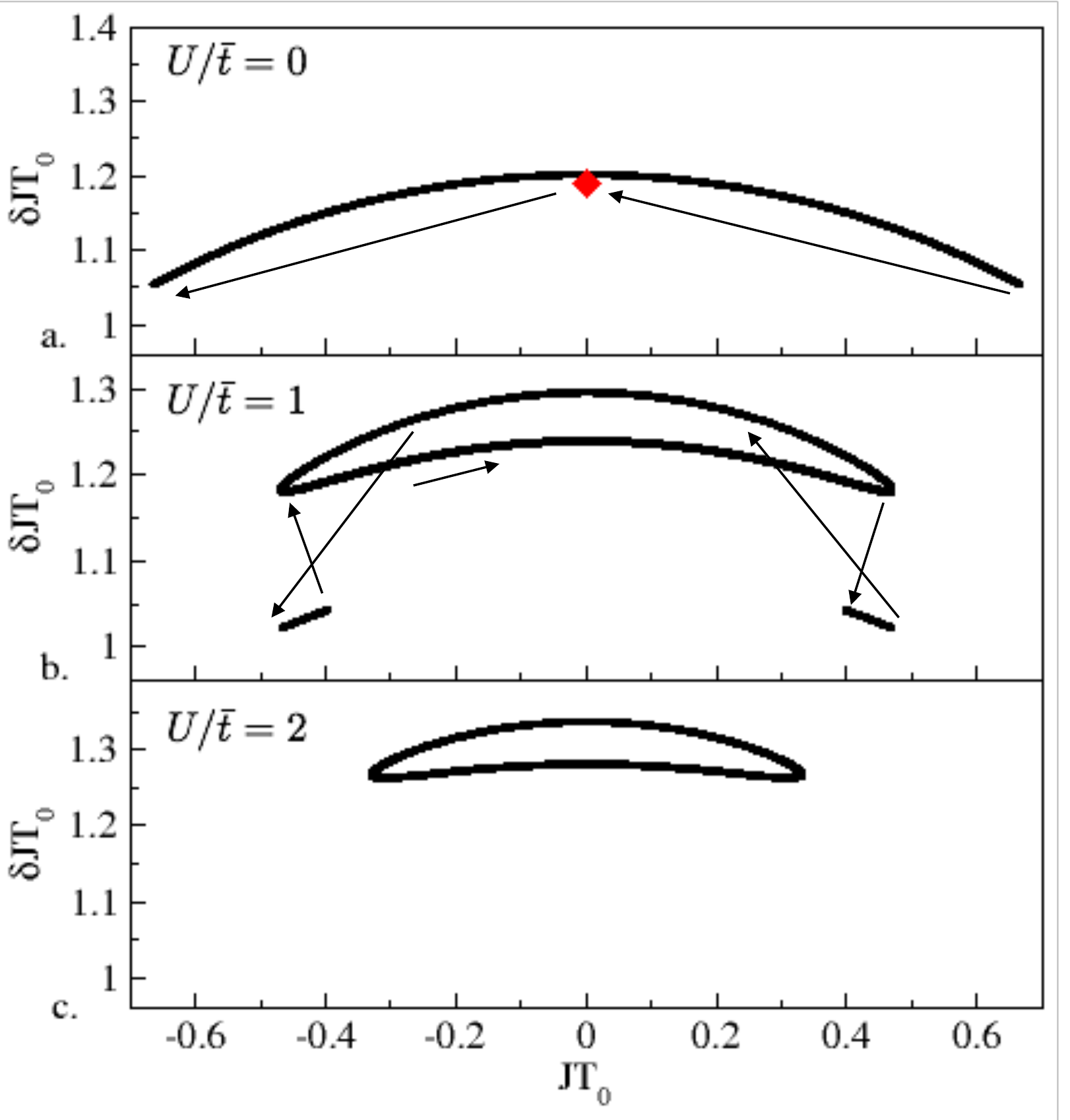}
		\caption{$\delta J$ vs. $J$ plot of a benzene-like lattice with both NN and NNN hopping. Here $\bar{t}' = \bar{t}/2$ and we focus on the range where $\pi/2 \leq \phi \leq 3\pi/2$. The interaction strengths are (a) $U/\bar{t} = 0$, (b) $U/\bar{t} = 1$, and (c) $U/\bar{t} = 2$. }
		\label{Fig:Hyst_NNN2}
	\end{figure}

	\subsection{Effects of next-nearest-neighbor hopping}
	In the presence of weak NNN hopping, the results resemble the case with only NN hopping. This applies to the noninteracting as well as the interacting cases. Fig~\ref{Fig:Curr&Hyst_NNN5} shows the current and its variance for a system with both NN and NNN hopping and $\bar{t}' = \bar{t}/5$. The resulting Lissajous curves of $\delta J$ vs. $J$ still exhibit the same structures as the case without the NNN hopping.
	
	However, adding a strong NNN hopping term can change the energy spectrum and the  periodicity of the $\phi$ dependence. Interestingly, additional level crossings appear even in the presence of weak interactions. The new level crossings lead to discontinuities in the current and current variance as shown in Fig~\ref{Fig:Curr&Hyst_NNN2} for the case with $\bar{t}' = \bar{t}/2$. Level crossings induced by interactions have been calculated using the Bethe Ansatz solution for the persistent current in a one-dimensional Hubbard model with strong interactions \cite{HubbardPC_BetheSoln, QuantumRingsForBeginners}. Appearance of interaction-induced level crossings, in violation of the non-crossing rule, results from non-trivial symmetries and underlying conservation laws (see Ref.~\cite{LiebBook, HubbardDegen} for a detailed mathematical discussion). The Lissajous curves break at the level crossing as shown in Fig~\ref{Fig:Hyst_NNN2}. If the range of the Peierls phase is restricted to $5\pi/6 \leq \phi \leq 7\pi/6$, there are no degeneracies and the Lissajous curve without the breaks can be observed.
	
	
	\begin{figure}
		\includegraphics[width=3.5in]{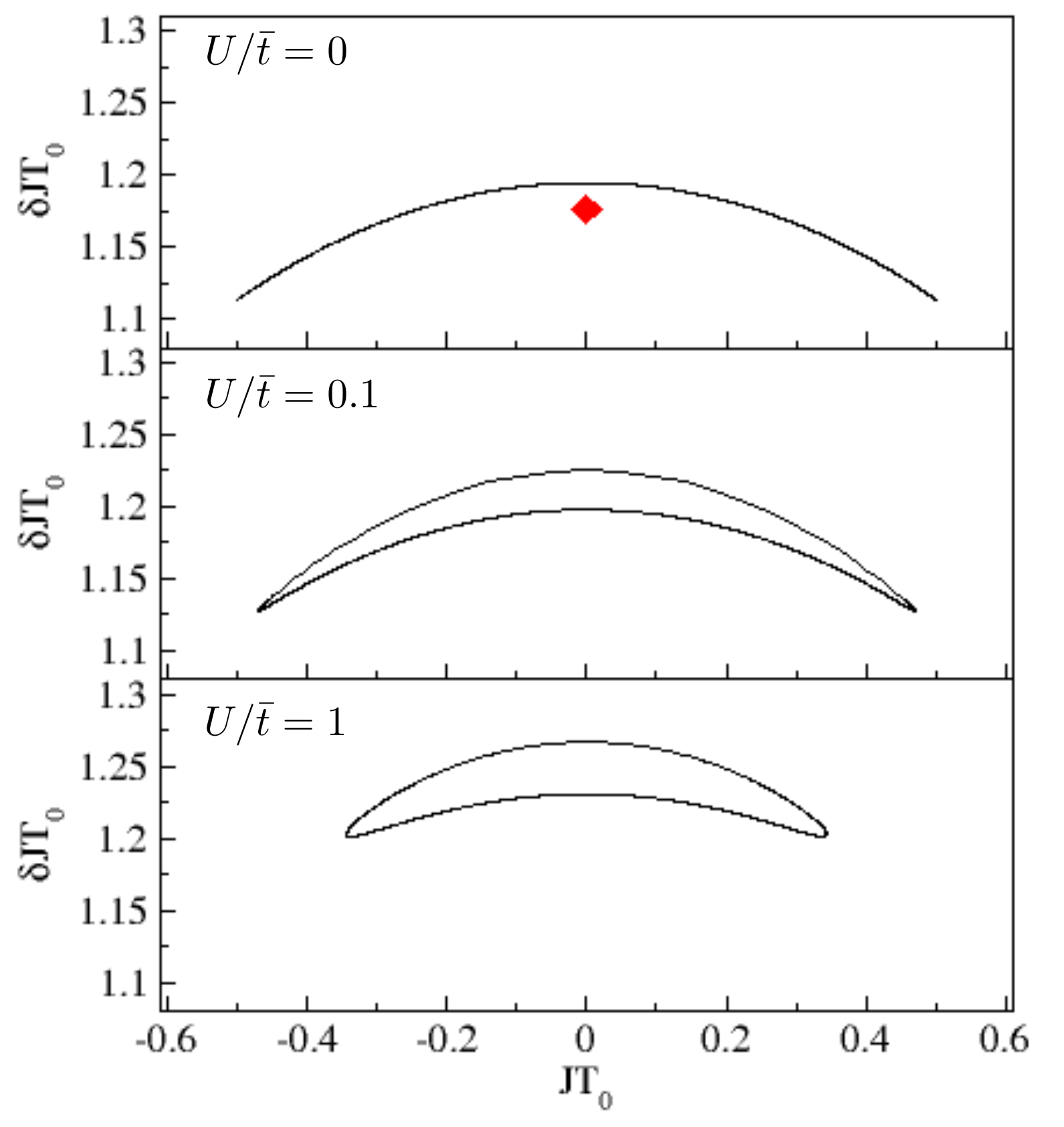}
		\caption{$\delta J$ vs. $J$ plot of an octagonal, half-filled lattice and we focus on the range where $\pi/2 \leq \phi \leq 3\pi/4$. The interaction strengths are (a) $U/\bar{t} = 0$, (b) $U/\bar{t} = 0.1$, and (c) $U/\bar{t} = 1$.}
		\label{Fig:Hyst_L8}
	\end{figure}
	
	\begin{figure}
		\includegraphics[width=3.5in]{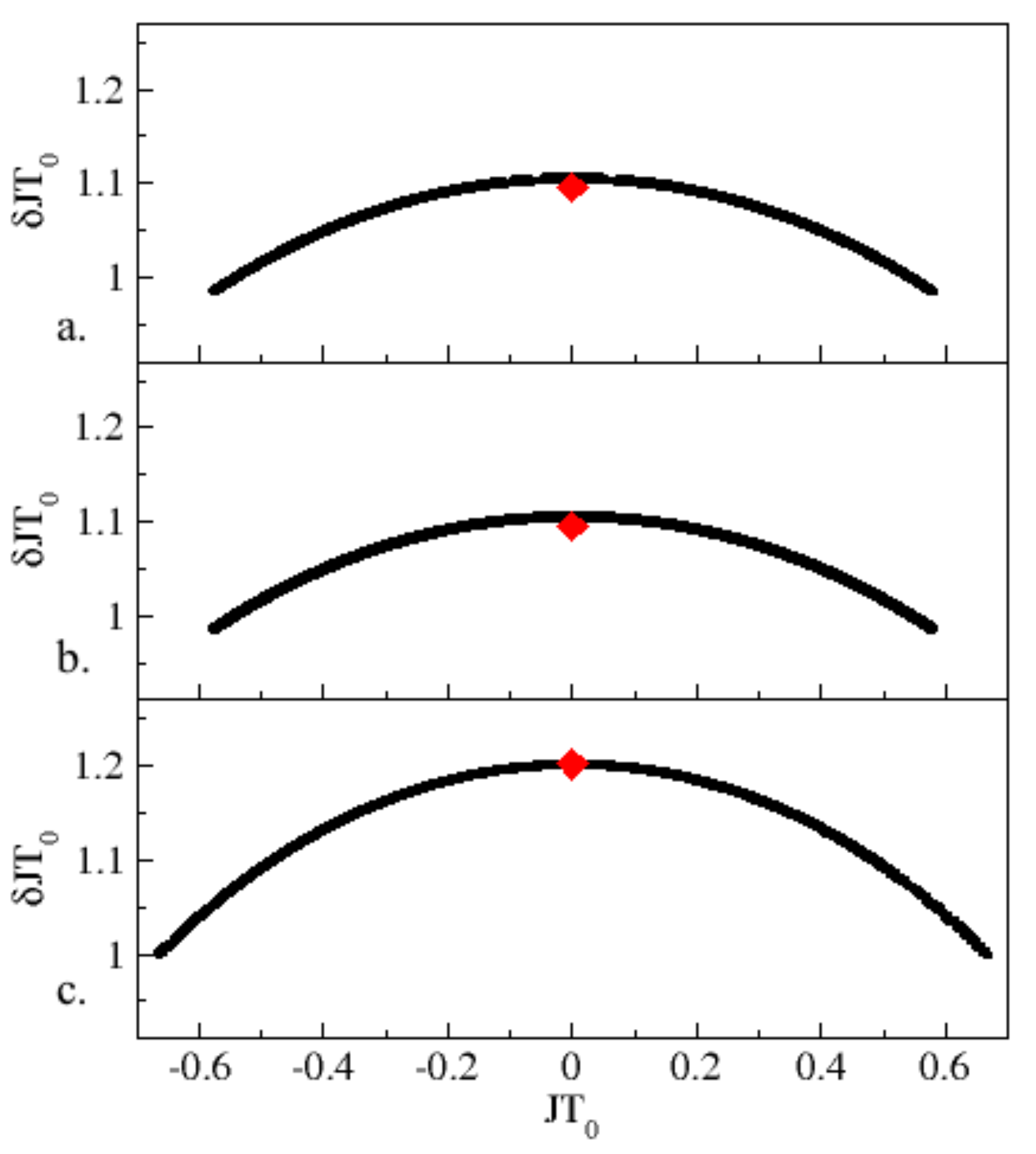}
		\caption{$\delta J$ vs. $J$ plot of the benzene-like lattice with $U/\bar{t} = 0$ for (a) $N_\uparrow = N_\downarrow = 2$, (b)$N_\uparrow = 2$, $N_\downarrow = 4$ in the range $2\pi/3 \leq \phi \leq \pi$, and (c) $N_\uparrow = 3$, $N_\downarrow = 0$ in the range $\pi/2 \leq \phi \leq 5\pi/6$. The red rhombuses show the discontinuities due to level crossings.}
		\label{Fig:Hyst_VarFill}
	\end{figure}
	
	\subsection{Effect of Lattice Size and Particle Filling}
	Lissajous curves are not only present for a hexagonal ring, but are present in rings with a larger number of sites. We tested a closed system with $L=8$ sites and found the interaction-induced loops are still present, as shown in Figure~\ref{Fig:Hyst_L8}. However, as the system approaches the thermodynamic limit the continuous energy band will replace the discrete energy levels. The origin of the Lissajous curves stems from the degenerate points in the non-interacting system and the avoided crossings in the interacting case. 
	
	One-to-one correspondence of current and its variance is not special to systems with half-filling. To demonstrate the presence of Lissajous curves away from half-filling, we tested even particle number, a partially polarized system, and a fully polarized system in the benzene-like lattice configuration with absence of interactions as shown in Figure~\ref{Fig:Hyst_VarFill}. Single-valued Lissajous curves with a disconnected point due to degeneracies are observable for all particle fillings, meaning the observed results are independent of particle configurations. 
	
	We remark that the degenerate points of the noninteracting system are non-analytic in the sense that the zero-temperature limit ($\mathcal{T}\rightarrow 0$) and the non-interacting limit $U\rightarrow 0$ are not compatible. To see this, we assume the Peierls phase is tuned to the degenerate point (say, $\phi=\pi/2$). If $U=0$ and $\mathcal{T}\rightarrow 0$, all the degenerate states will have the same statistical weight because the weight only depends on the energy of a state. Therefore, all the degenerate states contribute to any physical quantity in the $U=0$, $T\rightarrow 0$ limit. In contrast, we consider the other limit with $\mathcal{T}=0$ and $U\rightarrow 0$. No matter how small $U$ is, it serves as a perturbation to the hopping Hamiltonian and resolves the degeneracy. As a consequence, all physical quantities are from expectation values with respect to the genuine many-body ground state. Therefore, the two limits introduce different ways of obtaining physical quantities. Since cooling a system to $\mathcal{T}=0$ is prohibited by the third law of thermodynamics~\cite{Schroeder_book}, here we will follow the $U=0$, $\mathcal{T}\rightarrow 0$ limit to investigate the noninteracting system at the degeneracy point.
	
	\begin{figure}
		\includegraphics[width=3.5in]{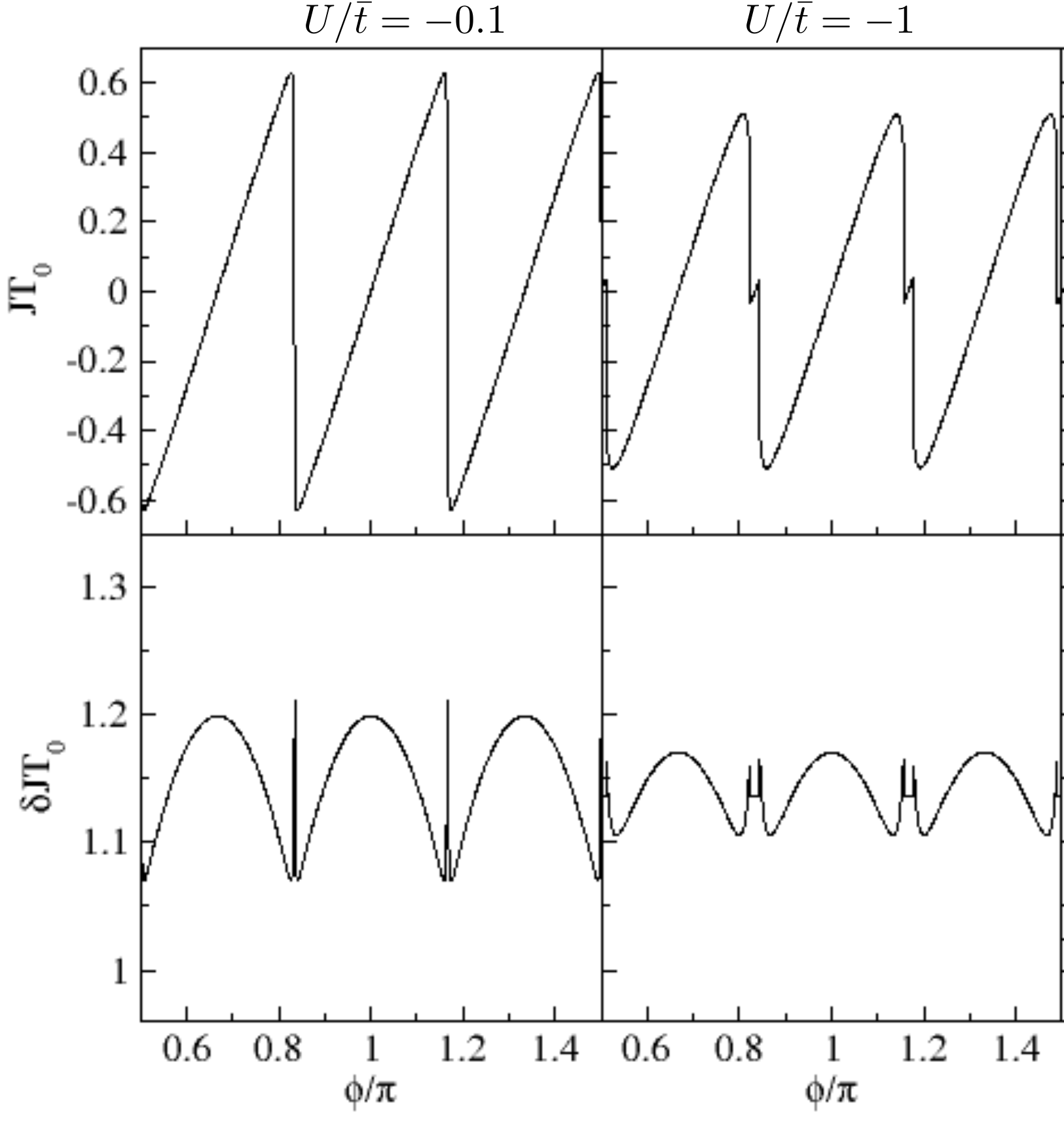}
		\caption{(Upper panels) Ground-state persistent currents $J$ and (Bottom panels) the standard deviation of current $\delta J$ as functions of the Peierls phase $\phi$ for the benzene-like lattice at half filling with (left column ) $U/\bar{t} = -0.1$ and (right column) $U/\bar{t} = -1$. }
		\label{Fig:CurVar_NegU}
	\end{figure}
	
	\begin{figure}
		\includegraphics[width=3.5in]{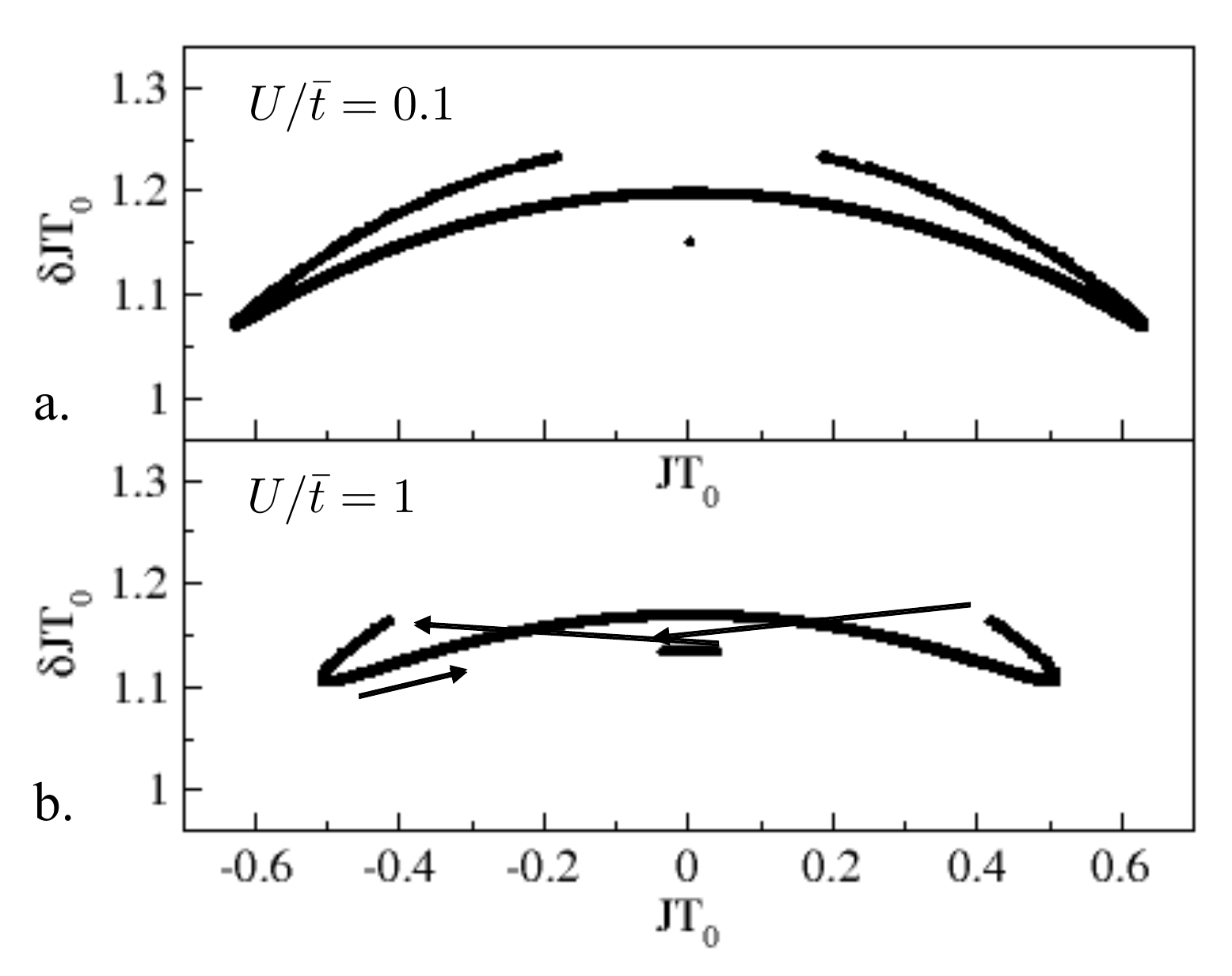}
		\caption{$\delta J$ vs. $J$ plot of a half-filled benzene-like lattice with nearest-neighbor hopping and attractive interaction. Here (a) $U/\bar{t} = -0.1$ and (b) $U/\bar{t} = -1$. The arrows show how the system evolves as the Peierls phase $\phi$ increases.}
		\label{Fig:Hyst_NegU}
	\end{figure}
	

	\subsection{Attractive interactions}
	We have also considered the effects of attractive interactions, $U < 0$, between fermions of opposite spin in the weakly and strongly interacting regimes. Adding weak attractive interactions induces a level crossing which is further accentuated as the interaction strength increases as shown in  Figure~\ref{Fig:CurVar_NegU}. The level crossing causes a double peak in the ground state current fluctuations. Plotting $\delta J$ vs. $J$ the Lissajous curve remains multivalued, but has a discontinuity corresponding to the level crossing as shown in Figure~\ref{Fig:Hyst_NegU} (a,b). This behavior is similar to the addition of significant next-nearest neighbor hopping to the repulsive interaction case (like the $\bar{t}^\prime = \bar{t}/2$ case shown previously), where an interaction induced level crossing results in discontinuities in $J$ and $\delta J$.

	\section{Additional information for open systems}
	\subsection{Details of the open-system approach}
	The Lindblad equation (7) shown in the main text was derived with the following assumptions.
	First, the coupling between the system and environments should be weak so that in the Born approximation the frequency scale of the system-reservoir coupling is small compared to the dynamical frequency scales of the system and reservoirs themselves.
	Second, the Markovian approximation requires the system-reservoir coupling to be time-independent over a short time scale, and the environment can rapidly return to equilibrium without being altered by the coupling.
	Moreover, there should be no memory in the reservoirs~\cite{breuer2007theory,weiss2012quantum}.
	
	In the Lindblad equation, we can express the density matrix in terms of Fock basis and simulate the dynamics by the fourth order Runge-Kutta algorithm~\cite{NRE}.
	The time evolution is simulated with adaptive time step as small as $\delta t \!=\!10^{-3}T_0$ and terminated when the steady state is reached.
	As the reservoir-system coupling, $\gamma$, becomes very large the simulation becomes unstable and smaller time steps are required.
	
	\subsection{Double-site lattice}
	\begin{figure}[t]
		\begin{center}
			\includegraphics[width=\columnwidth]{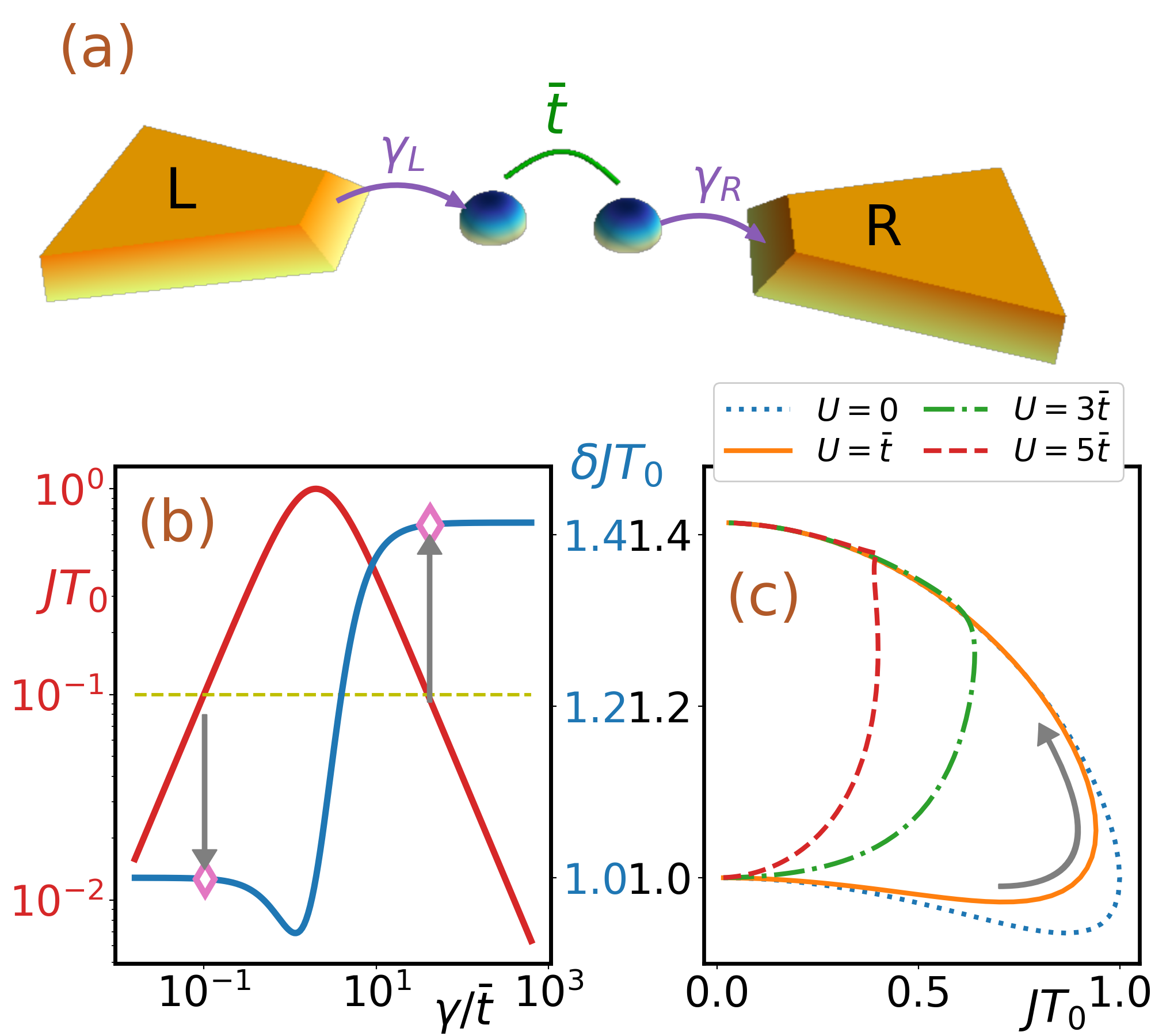}
			\caption{
				(a) A double-site lattice is connected to the source (left) and drain (right) as an open quantum system where the hopping coefficient between the sites is $\overline{t}$ and the on-site interaction has strength $U$.
				(b) The current $J$ (red) and standard deviation of the current $\delta J$ (blue) through the double-site lattice as functions of $\gamma$ in the noninteracting case.
				The asymptotic values match Eqs.~\eqref{eq:oqs_j} and~\eqref{eq:oqs_dj}.
				The yellow dashed line shows that there are two values of $\gamma$ giving rise to the same current $JT_0\!=\!0.1$ but with different current standard deviation (empty diamonds).
				(c) The current standard deviation versus average current in the double-site open quantum systems with selected values of $U$. The arrow indicates the direction of increasing $\gamma$.
			}
			\label{fig:OQS_2QDs_U}
		\end{center}
	\end{figure}
	
	To investigate the current variance in open systems, we first consider a double-site lattice system connected to two reservoirs as illustrated in Fig.~\ref{fig:OQS_2QDs_U}(a).
	The Hamiltonian of the double-site lattice is
	\begin{equation}\label{eq:H2qds}
	\mathcal{\hat H}_{\text{lat}}=-\overline{t}\sum_{\langle ij\rangle,\sigma}(\hat c^\dagger_{i\sigma}\hat c_{j\sigma}+h.c.)+\sum_{i=1}^{2}U \hat n_{i\uparrow}\hat n_{i\downarrow},
	\end{equation}
	where $\hat c^\dagger_n$ ($\hat c_n$) is the fermion creation (annihilation) operator on site $n$ and the density operator $\hat n_{i\sigma}\!=\!\hat c^\dagger_{i\sigma}\hat c_{i\sigma}$.
	Here we take an open-system approach by considering the lattice connected to two environments with coupling $\gamma_L$ and $\gamma_R$ respectively.
	The left (right) environment acts as a particle source (drain) which pumps (takes) particles into (out of) the lattice.
	
	The dynamics in the open-system approach is modeled by the Lindblad equation~\cite{breuer2007theory,DiVentra:2010ks,weiss2012quantum,Lai:2018vx} shown in the main text.
	Here, we choose $\gamma_L\!=\!\gamma_R\!=\!\gamma$ and use $\bar{t}$ as the energy unit.
	The time unit is $T_0\!=\!\hbar/\bar{t}$.
	We are interesting in the steady state where $d\rho/dT\!=\!0$ in the long time limit $T\!\rightarrow\!\infty$, and a steady state current can be identified.
	
	In the absence of interaction, $U\!=\!0$, the open-system Lindblad equation can be rewritten in terms of the single-particle correlation matrix which is demonstrated in Ref.~\cite{Lai:2018vx} and the equation of motion can be solved {\it exactly}.
	The steady-state current on the link between the adjacent sites is associated with the off-diagonal element of the single-particle correlated matrix.
	Explicitly, the current through the double-site lattice is
	\begin{equation}\label{eq:oqs_j}
	J=\frac{4\gamma \overline{t}}{4\overline{t}^2+\gamma^2}\approx
	\begin{cases}
	\gamma / \overline{t}&\gamma\ll \overline{t}\\
	4\overline{t} / \gamma&\gamma\gg \overline{t}.
	\end{cases}
	\end{equation}
	Here, the current shows $\gamma$ ($1/\gamma$) dependence in small (large) $\gamma$ regime and is symmetric around $\gamma\!=\!2\overline{t}$.
	Therefore, the current can be tuned through the coupling to the source/drain and the same current can be found corresponding to two different $\gamma$'s.
	The current standard deviation of noninteracting fermions can also be determined from Wick's theorem~\cite{altland2010condensed,mahan2000many}, and we arrive at
	\begin{equation}\label{eq:oqs_dj}
	\delta J =\sqrt{ \frac{2\overline{t}^2(8\overline{t}^4+2\overline{t}^2\gamma^2+\gamma^4)}{(4\overline{t}^2+\gamma^2)^2} }\approx
	\begin{cases}
	\overline{t} &\gamma\ll \overline{t}\\
	\sqrt{2}\overline{t} &\gamma\gg \overline{t}.
	\end{cases}
	\end{equation}
	Both the current and its standard deviation with different $\gamma$ are shown in Fig.~\ref{fig:OQS_2QDs_U}(b).
	More importantly, this immediately indicates that the current variance can be quantitatively different even if one operates in different $\gamma$'s and having the same current between the two sites.
	Figure~\ref{fig:OQS_2QDs_U}(c) shows that multivalues of $\delta J$ can be found for the same value of $J$, establishing the multi-valuedness between those two quantities. Therefore, the two-site system behaves differently from systems with three or more sites because both of the sites are directly connected to the reservoirs.

	In the presence of interactions, the current and current variance can be obtained by using the time-evolved density matrix described by the Lindblad equation.
	Figure~\ref{fig:OQS_2QDs_U}(c) shows the results for interacting fermions flowing through a two-site lattice. Although the on-site repulsion suppresses the current flowing through the lattice, the current and its standard deviation still show multi-valued behavior.
	It is worth mentioning that one may complete the loop in the $\delta J$ vs. $J$ plot by adding the curve from a similar setup with the source and drain reversed, i.e., by considering the current flowing in the reversed direction. However, the loop formed this way is not a Lissajous curve because the system does not traverse the whole curve in one setup.


\end{document}